\documentclass[aip,jap,reprint,twocolumn,superscriptaddress]{revtex4-1}
\usepackage{amsmath,amssymb,amsfonts,graphicx,bm}
\begin{document}

\title{Near-field thermodynamics: useful work, efficiency, and energy harvesting}

\author{Ivan Latella}
\email{ilatella@ffn.ub.edu}
\affiliation{Departament de F\'{i}sica Fonamental, Facultat de F\'{i}sica, Universitat de Barcelona, Mart\'{i} i Franqu\`{e}s 1, 08028 Barcelona, Spain}

\author{Agust\'in P\'erez-Madrid}
\email{agustiperezmadrid@ub.edu}
\affiliation{Departament de F\'{i}sica Fonamental, Facultat de F\'{i}sica, Universitat de Barcelona, Mart\'{i} i Franqu\`{e}s 1, 08028 Barcelona, Spain}

\author{Luciano C. Lapas}
\email{luciano.lapas@pq.cnpq.br}
\affiliation{Universidade Federal da Integra\c{c}\~{a}o Latino-Americana, Caixa Postal 2067, 85867-970 Foz do Igua\c{c}u, Brazil}

\author{J. Miguel Rubi}
\email{mrubi@ub.edu}
\affiliation{Departament de F\'{i}sica Fonamental, Facultat de F\'{i}sica, Universitat de Barcelona, Mart\'{i} i Franqu\`{e}s 1, 08028 Barcelona, Spain}
\affiliation{Department of Chemistry, Imperial College London, SW7 2AZ London, United Kingdom
}


\begin{abstract}
We show that the maximum work that can be obtained from the thermal radiation emitted between two planar sources in the near-field regime is much larger than that corresponding to the blackbody limit. This quantity as well as an upper bound for the efficiency of the process are computed from the formulation of thermodynamics in the near-field regime. The case when the difference of temperatures of the hot source and the environment is small---relevant for energy harvesting---is studied in detail. We also show that thermal radiation energy conversion can be more efficient in the near-field regime. These results open new possibilities for the design of energy converters that can be used to harvest energy from sources of moderate temperature at the nanoscale.
\end{abstract}


\newcommand{\dif}{d}
\newcommand{\ee}{e}
\newcommand{\ii}{i}
\newcommand{\kB}{k_\text{B}}
\newcommand{\sub}[1]{\text{#1}}
\newcommand{\vect}[1]{\bm{#1}}

\maketitle

\section{Introduction}

As reported in recent years, radiative heat transfer is enhanced by several orders of magnitude in the near-field as compared to the blackbody limit~\cite{Rousseau:2009,Shen:2009,Shen:2012}, thus increasing the amount of energy exchanged between bodies separated by a submicron distance~\cite{Perez-Madrid:2008,Perez-Madrid:2009,Perez-Madrid:2013}.
At the interface of polar materials, the coupling of phononic excitations with the electromagnetic fields results in the so-called surface phonon-polaritons (SPP).
These surface waves can be thermally excited at the nanoscale due to their existence in the infrared~\cite{Mulet:2002,Joulain:2005}.
Moreover, the physical mechanism leading to the increase of contributing electromagnetic modes derives from the local density of states, which plays a key role in the determination of thermodynamic functions~\cite{Joulain:2003,Dorofeyev:2011,Narayanaswamy:2013}.
When two planar sources supporting SPP are placed at a distance smaller than the thermal wavelength, the resonance of these modes is responsible for the considerable increase of the emitted radiation~\cite{,Joulain:2005,Volokitin:2007}.
This enhancement in the radiative heat transfer has a huge potential for generating clean and renewable energy from thermal sources using thermophotovoltaic devices \cite{Laroche:2006,Park:2008,Messina:2013}.

In the context of thermal radiation energy conversion, however, little attention has been paid to situations where the difference of temperatures of the sources is not too high.
The main reason is that the efficiency of the converters and the available energy to produce work is small in these cases.
Near-field radiation is a promising option to overcome this obstacle since, as we shall see, both efficiency and available energy can be higher than that of conventional blackbody radiation.
This mechanism opens the possibility to implement converters to harvest energy from sources of moderate temperature at the nanoscale.
For designing energy-conversion devices it is then crucial to estimate the maximum amount of work available from near-field radiation.

Our aim here is to compute the work flux that can be extracted from the radiation emitted by sources supporting SPP in the near-field regime.
We also obtain and discuss an upper bound for the efficiency of this process.

\section{Thermodynamics of thermal radiation}

Consider the energy flux $\dot{U}(T)$ radiated per unit time and surface by a body at absolute temperature $T$.
This energy flux can be written as
\begin{equation}
\dot{U}(T)=\int_0^\infty\dif \omega\; \hbar\omega n(\omega,T)\varphi(\omega),
\label{energy_flux}
\end{equation}
where $n(\omega, T)=\left(\ee^{\hbar\omega/\kB T}-1\right)^{-1}$ is the average number of photons in a single mode of frequency $\omega$, and $\hbar$ and $\kB $ are Planck's and Boltzmann's constants, respectively. The function $\varphi(\omega)$ in (\ref{energy_flux}), which we call the spectral flux of modes, will be obtained in the next section from the fluctuating electrodynamics approach \cite{Polder:1971}.

Since $T$ is an absolute temperature, the entropy flux $\dot{S}(T)$ associated to the radiation must satisfy the relation
\begin{equation}
\frac{1}{T}=\frac{\dif \dot{S}}{\dif \dot{U}}.
\end{equation}
Hence, the entropy flux is readily obtained from $\dot{U}(T)$ and is given by 
\begin{equation}
\dot{S}(T)=\int_0^T \dif T'\frac{1}{T'}\frac{\dif \dot{U}(T')}{\dif T'}. 
\label{ent:1}
\end{equation}
By introducing
\begin{equation}
\begin{split}
m(\omega,T)&= \left[1+n(\omega,T)\right]\ln\left[1+n(\omega,T)\right] \\
&\qquad-n(\omega,T)\ln n(\omega,T),  
\end{split}
\end{equation} 
the entropy flux (\ref{ent:1}) can be rewritten as
\begin{equation}
\dot{S}(T)=\int_0^\infty\dif \omega\; \kB  m(\omega,T) \varphi(\omega).
\label{Entropy}
\end{equation}
The expression (\ref{Entropy}) for the entropy flux is valid provided the spectral flux of modes does not depend on temperature.

In the situation at hand, we are interested in the case where there is interaction with a second body that emits thermal radiation at a different temperature. 
Thus, we will assume that each body always remains at the same fixed temperature, so that each emission spectrum is characterized by its own equilibrium temperature.
The fluxes associated only to the radiation of one of these bodies can be obtained by considering that the other body has zero temperature, but its optical properties such as absorption or reflection of incoming electromagnetic radiation have to be taken into account.
Here we also assume that both bodies have a planar surface.

\begin{figure}
\includegraphics[scale=0.9]{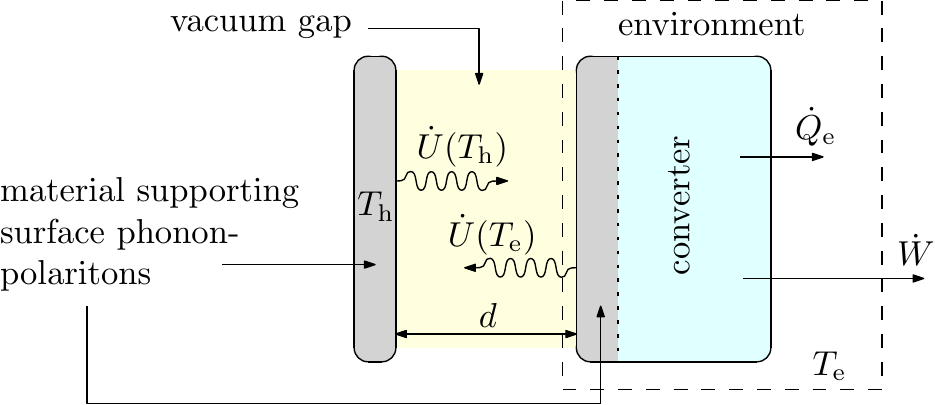} 
\caption{(color online) Schematic representation of the system.
The radiating surfaces are separated by a distance $d$ and a converter is coupled to the body that is in thermal equilibrium with the environment.
\label{sketch}
}
\end{figure}

In order to define the thermodynamic scheme of the conversion process, let us consider the radiation emitted by the surface of one of these materials at temperature $T_\sub{h}$ to the second radiating surface at environmental temperature $T_\sub{e}$, assuming $T_\sub{e}<T_\sub{h}$.
Now consider a converter that transforms the energy flux of the radiation incoming on the surface of the body at $T_\sub{e}$, delivering a certain amount of work flux $\dot{W}$.
We do not need to specify how the converter works because here we focus on an upper bound for the efficiency, as will become clear below.

The converter can be thought of as coupled to the body that is in thermal equilibrium with the environment, as sketched in~FIG.~\ref{sketch}.
Thus, the energy flux balance equation for the system plus the environment can be written as
\begin{equation}
\Delta \dot{U}+ \dot{Q}_\sub{e}+\dot{W}=0,
\label{first_law}
\end{equation}
where $\dot{Q}_\sub{e}$ is the heat flux delivered to the environment, and 
\begin{equation}
\Delta \dot{U}=\dot{U}(T_\sub{e})-\dot{U}(T_\sub{h}) 
\end{equation}
is the variation of energy flux of the radiation.
Furthermore, according to the formulation of the second law of thermodynamics~\cite{deGroot:1984}, one has
\begin{equation}
\Delta \dot{S}+\Delta \dot{S}_\sub{e}=\Delta \dot{S}_\sub{irr}\geq0, 
\label{second_law}
\end{equation}
where $\Delta \dot{S}_\sub{irr}$ is the entropy production flux due to irreversibilities in the processes of conversion,  $\Delta \dot{S}_\sub{e}$ is the variation of entropy flux of the environment, and
\begin{equation}
\Delta \dot{S}=\dot{S}(T_\sub{e})-\dot{S}(T_\sub{h}) 
\end{equation}
is the variation of entropy flux of the radiation.
Here we have assumed that in the converter there are no sources either of energy or entropy, i.e., stationary regime.
Moreover, equations (\ref{first_law}) and (\ref{second_law}) are closely linked because $\Delta\dot{S}_\sub{e}=\dot{Q}_\sub{e}/T_\sub{e}$.
Under these conditions, the work flux provided by the device reads 
\begin{equation}
\dot{W}=T_\sub{e}\Delta \dot{S}-\Delta \dot{U}-T_\sub{e}\Delta \dot{S}_\sub{irr}, 
\end{equation}
and in the limiting case when there is no entropy production, one has an ideal work flux 
\begin{equation}
\dot{\mathcal{W}}\equiv T_\sub{e}\Delta \dot{S}-\Delta \dot{U}. 
\end{equation}
Therefore, $\dot{\mathcal{W}}$ is the maximum work flux that can be obtained in the process of conversion of the incoming energy flux.

The efficiency $\eta$ is given by the ratio of the available work flux to the input energy flux.
According to the previous scheme, the latter here is the energy flux coming from the hot source $\dot{U}(T_\sub{h})$, whence one has~\cite{Kjelstrup:2010}
\begin{equation}
\eta\equiv\frac{\dot{W}}{\dot{U}(T_\sub{h})}= \frac{\dot{\mathcal{W}}-T_\sub{e}\Delta \dot{S}_\sub{irr}}{\dot{U }(T_\sub{h})}.
\label{first_law_efficiency}
\end{equation}
An upper bound $\bar{\eta}$ for the efficiency is obtained by considering that the work flux is ideal; i.e., 
\begin{equation}
\bar{\eta}=\frac{\dot{\mathcal{W}}}{\dot{U}(T_\sub{h})}\geq \eta. 
\end{equation}
In Sec.~\ref{efficiency:bounds} we obtain this upper bound $\bar{\eta}$ as well as $\dot{\mathcal{W}}$ considering near-field thermal radiation.

\section{Spectral flux of modes between two media}

According to the scheme discussed in the previous section, here we consider two isotropic semi-infinite nonmagnetic media separated by a vacuum gap.
In order to determine the spectral flux of modes $\varphi(\omega)$, first it is assumed that the temperature of one of these media, say medium~1, is $T_1$ and the temperature of medium~2 is $T_2=0$.
Moreover, thermal excitations in medium~1 produce fluctuating currents that generate fluctuating electromagnetic fields in the vacuum gap.
Thus, denoting by $z$ the direction perpendicular to the surfaces, $\vect{e}_z$ the unit vector in this direction, and $\vect{E}_1$ and $\vect{H}_1$ the fluctuating electromagnetic fields due to currents in medium~1, the energy flux radiated by medium~1 and absorbed by medium~2 is given by~\cite{Mulet:2002,Biehs:2012}
\begin{equation}
\dot{U}(T_1)=\left\langle \Sigma_z^{1\to2}\right\rangle=\left\langle \vect{E}_1\times \vect{H}_1\right\rangle\cdot \vect{e}_z.
\end{equation}
In the previous equation $\Sigma_z^{1\to2}$ is the normal component of the Poynting vector and $\langle\,\cdots\rangle$ denotes statistical average. Analogously, computing the energy radiated by medium~2 and absorbed by medium~1, $\dot{U}(T_2)=\left\langle \Sigma_z^{2\to1}\right\rangle$, the net energy transfer $\Delta\dot{U}=\dot{U}(T_2)-\dot{U}(T_1)$ takes the form
\begin{equation}
\Delta\dot{U}=\int_0^\infty\dif \omega\; \hbar\omega \left[n(\omega,T_2)-n(\omega,T_1)\right]\varphi(\omega),
\end{equation}
with $\varphi(\omega)$ for two identical media given by~\cite{Volokitin:2001,Mulet:2002,Joulain:2005,Volokitin:2007}
\begin{equation}
\begin{split}
\varphi(\omega)&=\sum_{\alpha=\mathrm{p},\mathrm{s}}\left\{\int_0^{\omega/c}\frac{\dif \kappa\, \kappa}{4\pi^2}\frac{\left[1-|R_\alpha(\kappa,\omega)|^2\right]^2}{\left|1-\ee^{2\ii\gamma d}R_\alpha^2(\kappa,\omega)\right|^2}\right.\\
&\left.\qquad\ +\int_{\omega/c}^\infty\frac{\dif \kappa\, \kappa}{\pi^2}\frac{\ee^{-2|\gamma|d}\mathrm{Im}^2\left[ R_\alpha(\kappa,\omega)\right]}{\left|1-\ee^{-2|\gamma|d}R_\alpha^2(\kappa,\omega)\right|^2}\right\}.
\label{SFM}
\end{split}
\end{equation}
Here $d$ is the width of the vacuum gap, $R_\alpha(\kappa,\omega)$ is the reflection coefficient of the vacuum-material interface for polarizations $\alpha=\mathrm{p},\mathrm{s}$, and $\kappa$ is the component of the wave vector parallel to the surfaces which is related to $\gamma$ through $\gamma=\sqrt{(\omega/c)^2-\kappa^2}$.
The case of blackbody radiation is obtained by assuming that the materials are perfect absorbers so that $R_\alpha=0$ and hence
\begin{equation}
\varphi(\omega)=\varphi_\sub{bb}(\omega)=\left(\frac{\omega}{2\pi c}\right)^2, 
\end{equation}
where the subscript bb refers to the blackbody regime.
Thus, in this case one obtains $\dot{U}=\dot{U}_\sub{bb}=\sigma T^4$ and $\dot{S}=\dot{S}_\sub{bb}=4\sigma T^3/3$, where $\sigma$ is Stefan's constant.
For the case of blackbody radiation, the ideal work flux is given by $\dot{\mathcal{W}}=\dot{\mathcal{W}}_\sub{bb}$, with
\begin{equation}
\dot{\mathcal{W}}_\sub{bb}=\sigma\left(T_\sub{h}^4-T_\sub{e}^4\right)-\frac{4}{3}\sigma T_\sub{e}\left(T_\sub{h}^3-T_\sub{e}^3\right).
\end{equation}

As noted previously, for gap sizes $d\ll \lambda_T=c\hbar/\kB T$ the emission is mainly dominated by SPP if the material supports them ($\lambda_T=7.6\,\mu$m for $T=300\,$K).
The enhancement of the radiative heat transfer is even more pronounced if both surfaces are identical because in this case these surface modes are resonantly excited~\cite{Mulet:2002}.
For this reason, we restrict here to sources made of the same material.
As examples of materials that support SPP, we will explicitly consider silicon carbide (SiC) and hexagonal boron nitride (hBN), whose optical data are taken from Ref.~\onlinecite{Palik:1998} for the former and from Ref.~\onlinecite{Messina:2013} for the latter.
However, we will present results in terms of the resonant frequency for any material that supports these surface modes.
The dielectric constants of SiC and hBN are suitably described by the Lorentz model
\begin{equation}
\varepsilon(\omega)=\varepsilon_\infty\frac{\omega^2_\sub{L}-\omega^2-\ii\Gamma\omega}{\omega^2_\sub{T}-\omega^2-\ii\Gamma\omega}, 
\end{equation}
where $\varepsilon_\infty$, $\omega_\sub{L}$, $\omega_\sub{T}$, and $\Gamma$ are material-dependent parameters. The dependence of $\varphi(\omega)$ on the dielectric constant enters through the reflection coefficients $R_\alpha(\kappa,\omega)$, as will be discussed below.

\section{Efficiency bounds and maximum work flux for near-field radiation emitted by polar materials}
\label{efficiency:bounds}

For polar materials and when the two surfaces are close enough, the spectral flux of modes in the near-field regime is mainly dominated by p-polarized evanescent modes~\cite{Joulain:2005,Volokitin:2007}.
This corresponds to the second term in curly brackets in (\ref{SFM}) with $\alpha=\mathrm{p}$ (s-polarized evanescent modes can be the dominant contribution for metals, see Ref.~\onlinecite{Chapuis:2008}). For this p-polarized radiation and in the electrostatic limit, the reflection coefficient does not depend on $\kappa$ and can be written as 
\begin{equation}
R_\mathrm{p}(\omega)=\frac{\varepsilon(\omega)-1}{\varepsilon(\omega)+1}. 
\end{equation}
Recently, interesting analytical results were found under these conditions in Ref.~\onlinecite{Rousseau:2012}.
There, the authors derived an approximate analytic closed-form expression for the heat transfer coefficient and studied its dependence on the temperature.
Our point is that the same arguments can be applied to compute the energy and entropy fluxes.
According to that method~\cite{Rousseau:2012}, introducing
\begin{equation}
f(\omega)=\frac{\mathrm{Im}\left[R_\mathrm{p}^2(\omega)\right]}{\mathrm{Im}^2\left[R_\mathrm{p}(\omega)\right]} 
\end{equation}
and $f'(\omega)=\dif f(\omega)/\dif \omega$,  we write the energy flux in the near-field regime as
\begin{align}
\dot{U}_\sub{nf}(T)&=\int_0^\infty \dif \omega\, \hbar\omega n(\omega,T) \frac{\mathrm{Im}\left[\mathrm{Li}_2(R_\mathrm{p}^2(\omega))\right]}{4\pi^2d^2f(\omega)}\label{integral_energy}\\
&\simeq \hbar\omega_0 n_0(T)\frac{\mathrm{Re}\left[\text{Li}_2\left(R^2_\mathrm{p}(\omega_0)\right)\right]}{4\pi d^2f'(\omega_0)}
\label{result_energy},
\end{align}
where the subscript nf stands for near-field, $\text{Li}_2(z)$ is the dilogarithm function, and we have introduced $n_0(T)\equiv  n(\omega_0,T)$, with
$\omega_0$ being the frequency of the SPP of the single interface.
Here we have assumed that there is only one resonant mode, as is the case with SiC and hBN for which 
\begin{equation}
\omega_0=\left(\frac{\varepsilon_\infty\omega_\sub{L}^2+\omega_\sub{T}^2}{\varepsilon_\infty+1}\right)^{1/2} 
\end{equation}
and $f'(\omega_0)\simeq-4/\Gamma$.
Equation (\ref{result_energy}) shows that near-field radiation is highly monochromatic with dominant frequency $\omega_0$.
Moreover, this result also implies that in this near-monochromatic approximation the spectral flux of modes in the near-field regime becomes
\begin{equation}
\varphi_\sub{nf}(\omega)= g_d(\omega)\delta(\omega-\omega_0), 
\end{equation}
with
\begin{equation}
g_d(\omega)=\frac{\mathrm{Re}\left[\mathrm{Li}_2\left(R^2_\mathrm{p}(\omega)\right)\right]}{4\pi d^2f'(\omega)},
\end{equation}
for the radiation emitted by the polar materials under consideration and in the presence of a nearby second surface made of the same material.
The function $g_d(\omega)$, which is restricted to the frequency of the resonant mode due to the Dirac $\delta$, contains the information about the emissivity of the material and the typical dependence $1/d^2$ on the gap size in this regime~\cite{Mulet:2002,Rousseau:2012}.
In addition, this functional form of the spectral flux of modes allows us to easily compute the flux associated to other thermodynamic quantities.
The entropy flux of near-field radiation is therefore given by
\begin{equation}
\dot{S}_\sub{nf}(T)=\int_0^\infty \dif \omega\, \kB  m(\omega,T)\varphi_\sub{nf}(\omega)=\kB  m_0(T)g_d(\omega_0),
\label{result_entropy}
\end{equation}
where $m_0(T)\equiv m(\omega_0,T)$.
Hence, considering $\dot{U}(T)=\dot{U}_\sub{nf}(T)$ and $\dot{S}(T)=\dot{S}_\sub{nf}(T)$, the ideal work flux in the near-field regime takes the form
\begin{equation}
\begin{split}
\dot{\mathcal{W}}_\sub{nf}&=\hbar\omega_0g_d(\omega_0)\Bigg\{\frac{\kB T_\sub{e}}{\hbar\omega_0}\left[m_0(T_\sub{e})-m_0(T_\sub{h})\right]\\
&\qquad\qquad\qquad\qquad-\left[n_0(T_\sub{e})-n_0(T_\sub{h})\right]\Bigg\}. 
\end{split}
\end{equation}
As a result, the upper bound for the efficiency in this regime is given by
\begin{equation}
\bar{\eta}_\sub{nf}=1-\frac{n_0(T_\sub{e})}{n_0(T_\sub{h})}+\frac{\kB T_\sub{e}}{\hbar\omega_0}\frac{m_0(T_\sub{e})-m_0(T_\sub{h})}{n_0(T_\sub{h})},
\end{equation}
which corresponds to that of near-monochromatic radiation~\cite{Landsberg:1980}, as expected.
For fixed temperatures, the upper bound $\bar{\eta}_\sub{nf}$ increases as the resonant frequency increases, but its growth is limited by the Carnot efficiency since $\lim_{\omega_0\to\infty}\bar{\eta}_\sub{nf}=1-T_\sub{e}/T_\sub{h}$.

\begin{figure}
\includegraphics[scale=1]{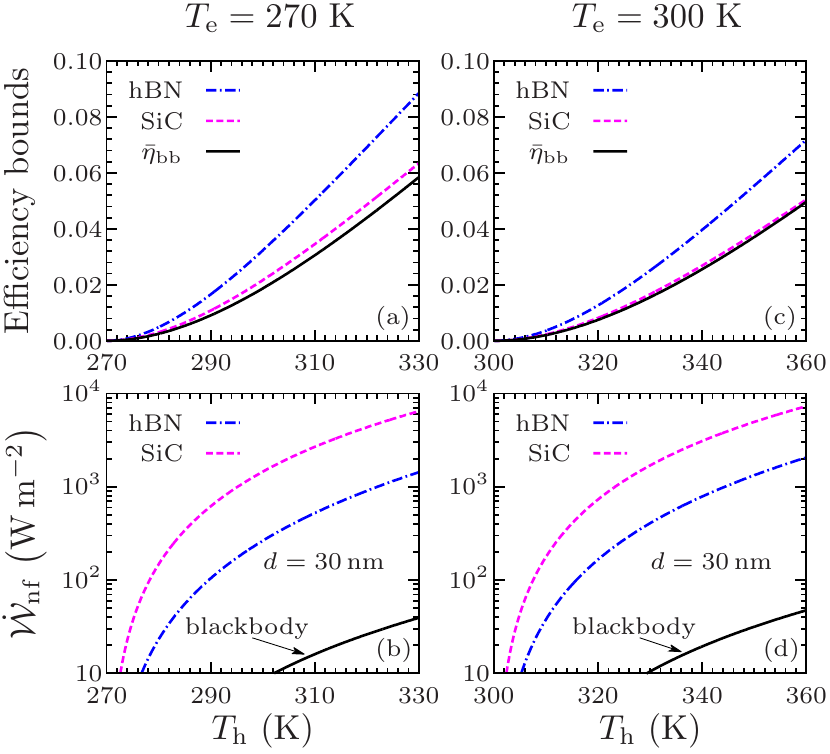} 
\caption{(color online) Efficiency bound $\bar{\eta}_\sub{nf}$ and ideal work flux $\dot{\mathcal{W}}_\sub{nf}$ as a function of the temperature of the hot source $T_\sub{h}$ for two different environmental temperatures $T_\sub{e}$.
In (a) and (b) these quantities are plotted for $T_\sub{e}=270$ K, while in (c) and (d) for $T_\sub{e}=300$ K. The quantities $\bar{\eta}_\sub{bb}$ and $\dot{\mathcal{W}}_\sub{bb}$ corresponding to blackbody radiation are also shown.
}
\label{graph_efficiency_work}
\end{figure}

\begin{figure}
\includegraphics[scale=1]{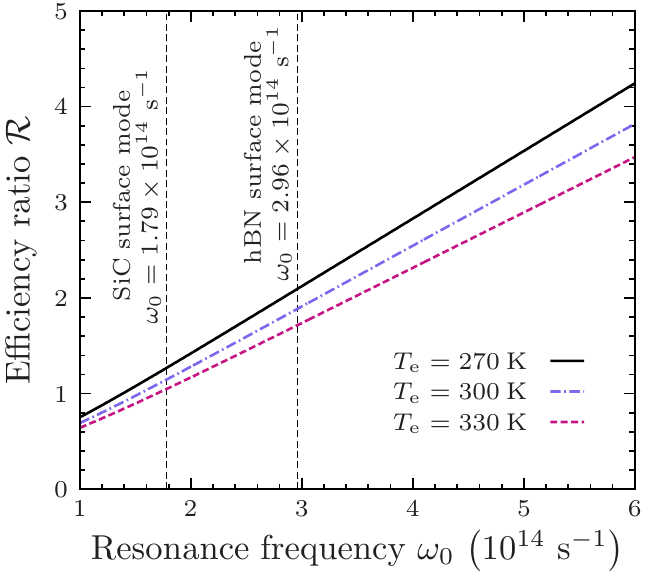} 
\caption{(color online) Ratio $\bar{\eta}_\sub{nf}$ to $\bar{\eta}_\sub{bb}$ in the limit where the temperature of the hot source approaches the environmental temperature as a function of the frequency of the SPP.
For $T_\sub{e}=270\,$K, in this limit $\bar{\eta}_\sub{nf}$ is more than twice as large as $\bar{\eta}_\sub{bb}$ for hBN.
\label{graph_R}
}
\end{figure}

The maximum work flux that can be extracted from the radiation in the near-field regime is considerably higher than that obtained from blackbody radiation, as shown in~FIG.~\ref{graph_efficiency_work} for SiC and hBN setting $d=30\,$nm, where also the bounds for the efficiency are plotted.
One therefore may wonder whether the bounds imposed by thermodynamics on the efficiency permit the process of conversion to be more efficient in the near-field regime as compared to the blackbody regime.
The upper bound for the efficiency in the blackbody limit is easily computed and given by~\cite{Landsberg:1980}
\begin{equation}
\bar{\eta}_\sub{bb}= 1-\frac{4}{3}\frac{T_\sub{e}}{T_\sub{h}}+\frac{1}{3}\left(\frac{T_\sub{e}}{T_\sub{h}}\right)^4. 
\end{equation}
Taking $\bar{\eta}_\sub{bb}$ as a reference provides a notion of how $\bar{\eta}_\sub{nf}$ varies for different values of the resonant frequency $\omega_0$.
For high enough temperatures, $\bar{\eta}_\sub{bb}$ can be higher than $\bar{\eta}_\sub{nf}$, however, here we concentrate on the case where the difference of temperatures of the surfaces is small in comparison with the average temperature.
Such a situation is physically relevant, for instance, if a converter is implemented in a certain device with the purpose of harvesting energy from near-field radiation, taking advantage of the fact that one of its components has a temperature somewhat higher than the environment due to some independent process.
The penalization in the efficiency of the conversion because of the small temperature difference can be compensated by the considerable amount of work flux obtained from near-field radiation.
Thus, we take the limit where the temperature of the hotter surface approaches the environmental temperature $T_\sub{e}$ in the ratio $\bar{\eta}_\sub{nf}/\bar{\eta}_\sub{bb}$ and obtain
\begin{equation}
\mathcal{R}\equiv \lim_{T_\sub{h}\to T_\sub{e}}\frac{\bar{\eta}_\sub{nf}}{\bar{\eta}_\sub{bb}}=\frac{\hbar\omega_0}{4\kB T_\sub{e}}\left[1-\exp\left(-\frac{\hbar\omega_0}{\kB T_\sub{e}}\right)\right]^{-1}.
\label{R}
\end{equation}
The condition $\bar{\eta}_\sub{nf}>\bar{\eta}_\sub{bb}$ in this limit, i.e. $\mathcal{R}>1$, can be numerically resolved and is satisfied if 
\begin{equation}
\omega_0>3.921 \frac{\kB T_\sub{e}}{\hbar}, 
\end{equation}
leading to a threshold frequency for which the conversion of near-field radiation can be more efficient than the conversion of blackbody radiation.
Furthermore, this also means that the higher the value of $\omega_0$, the more $\bar{\eta}_\sub{nf}$ increases at small temperature difference, see FIG.~\ref{graph_R}.
In contrast, the available work flux diminishes when the resonant frequency increases.
To see that, we approximate $\dot{\mathcal{W}}_\sub{nf}$ to leading order in $\Delta T=T_\sub{e}-T_\sub{h}$ such that $|\Delta T|/T_0\ll1$, with $T_0=(T_\sub{e}+T_\sub{h})/2$, and obtain
\begin{equation}
\dot{\mathcal{W}}_\sub{nf}\simeq\frac{g_d(\omega_0)}{8\kB T_0}\left[\hbar\omega_0\text{csch}\left(\frac{\hbar\omega_0}{2\kB T_0}\right)\frac{\Delta T}{T_0}\right]^2,
\label{work_expansion}
\end{equation}
which decreases for increasing $\omega_0$ because of the hyperbolic cosecant.
We see clearly that the choice of the material, characterized by $\omega_0$, directly affects the efficiency and determines the available work flux.

\section{Summary and conclusion}

In summary, we have formulated a thermodynamic scheme for the radiation in the near-field regime.
We have obtained analytical expressions for energy, entropy, and ideal work fluxes of near-field radiation using the near-monochromatic approximation~\cite{Rousseau:2012}.
The upper bound for the efficiency has been computed and has been shown to agree with previous results for near-monochromatic radiation~\cite{Landsberg:1980}.
We have elucidated the dependence of this bound on the resonance frequency and we have also seen how the choice of the material influences the performance of the energy conversion process.
This approach sheds light on thermal radiation energy conversion exploiting optical properties of the emitters and provides new perspectives for harvesting energy from near-field radiation.

\begin{acknowledgments}
This work was supported by the Spanish Government under Grant No. FIS2011-22603, and by CNPq and Funda\c{c}\~{a}o Arauc\'aria of the Brazilian Government.
IL acknowledges financial support through an FPI Scholarship (Grant No. BES-2012-054782) from the Spanish Government.
JMR acknowledges financial support from Generalitat de Catalunya under program ICREA Academia. 
\end{acknowledgments}

\bibliography{mainbib}

\end{document}